\documentclass[aps,pre,preprint,superscriptaddress]{revtex4-1}

\usepackage{amsmath,amssymb}
\usepackage{graphicx}
\usepackage{color}

\begin{document}

\preprint{}

\title{Real-space renormalization group for the transverse-field Ising model in two and three dimensions}


\author{Ryoji Miyazaki}
%\thanks{}
\author{Hidetoshi Nishimori}
\affiliation{Department of Physics, Tokyo Institute of Technology, Oh-okayama, Megro-ku, Tokyo 152-8551, Japan}
\author{Gerardo Ortiz}
\affiliation{Department of Physics, Indiana University, Bloomington, Indiana 47405, USA}


\date{\today}

\begin{abstract}
The two- and three-dimensional transverse-field Ising models with ferromagnetic
exchange  interactions are analyzed by means of the real-space renormalization group method.
The basic strategy is a generalization of a method developed for the one-dimensional case, which 
exploits the exact invariance of the model under renormalization and 
is known to give the exact values of the critical point and critical exponent $\nu$.
The resulting values of the critical exponent $\nu$ in two and three dimensions are in good agreement with those for the classical Ising model in three and four dimensions.
This is the first example in which a real-space renormalization group on ($2+1$)- and ($3+1$)-dimensional Bravais lattices yields accurate estimates of the critical exponents.
\end{abstract}

\pacs{05.10.Cc, 05.30.Rt, 05.50.+q, 64.60.F-}
%\keywords{}

\maketitle

\section{Introduction}

The real-space renormalization group framework has been developed some time ago and is often considered a crude approximation in practice because the truncation that is necessarily involved for tractability usually leads to unreliable estimates of critical exponents~\cite{Nishimori}.
Standard approaches on the basis of the block-spin transformation for quantum systems~\cite{Drell}-\cite{Fradkin} succeeded in describing qualitative properties but still had difficulties in quantitatively accurate calculations.
For the one-dimensional transverse-field Ising model, which is equivalent to the two-dimensional classical Ising model, Fernandez-Pacheco~\cite{Fernandez} modified the block construction of the standard block-spin transformation~\cite{Drell} to preserve the high symmetry of the model and could reproduce the exact values of the critical point and critical exponent $\nu$.
Although current activities in real-space renormalization group approaches to quantum systems are
often focused on numerically accurate evaluations, by using for instance the density-matrix renormalization 
group~\cite{White} or the multiscale entanglement renormalization ansatz~\cite{Vidal}, it is important to develop 
analytical or quasi-analytical methods applicable to higher dimensions since numerical methods are not always 
suitable for 
the calculation of critical properties.

The present paper reports on our successful generalization of the one-dimensional method of Fernandez-Pacheco~\cite{Fernandez} to two and three dimensions.
Although it is not possible to yield exact solutions for those higher-dimensional systems corresponding to three and four dimensions in the classical representation, at least the results for the critical exponent $\nu$ are impressive, given that quantitatively reliable real-space renormalization group methods in three- and four-dimensional classical models have not necessarily been established, see e.g.~\cite{Mazenko}-\cite{Fradkin} and \cite{Penson}-\cite{Mattis}.

In Section ~\ref{sec:1D} we explain the application of the real-space renormalization group method to the one-dimensional transverse-field Ising model and calculate the transition point and the critical exponents analytically.
The main results of the paper are presented in Sections ~\ref{sec:2D} and ~\ref{sec:3D}.
Here, we explain how to extend the real-space renormalization group method to higher spatial dimensions.
Our study is concluded in Section~\ref{sec:conclusions}.

\section{\label{sec:1D}Real-space renormalization group for the one-dimensional transverse-field Ising model}

Let us first review the one-dimensional scheme proposed in~\cite{Fernandez} and generalize it to the case with a longitudinal field,
\begin{equation}
H = -J \sum_{i=1}^{N-1} \sigma_{i}^{z} \sigma_{i+1}^{z} - h_{x} \sum_{i=1}^{N} \sigma_{i}^{x} - h_{z} \sum_{i=1}^{N} \sigma_{i}^{z}, \label{eq:1DIsingH}
\end{equation}
where free boundary conditions are assumed.
The system is composed of $S=1/2$ spins, and $\sigma_{i}^{z}$, $\sigma_{i}^{x}$ are the Pauli matrices.
We consider ferromagnetic interactions $J>0$ and assume that $h_{x}$ and $h_{z}$ are not negative without loss  of generality. This model is known to be exactly solvable when $h_z=0$, in which case the system is self-dual~\cite{Nishimori}.

\begin{figure}[b]
  \begin{center}
  \includegraphics[width=5cm,trim=0 250 0 0,clip]{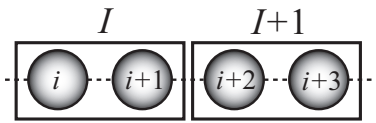}
  \caption{Construction of block spins in one dimension.}
  \label{fig:1Dblocks}
  \end{center}
\end{figure}

We start by dividing the lattice into blocks of two spins as shown in Fig.~\ref{fig:1Dblocks}.
The Hamiltonian is also split into the intra-block and the inter-block parts, 
\begin{gather}
H_{\mathrm{intra},I} = -J \sigma_{i}^{z} \sigma_{i+1}^{z} - h_{x} \sigma_{i}^{x} - h_{z} \sigma_{i}^{z}, \label{eq:Hintra1D} \\
H_{\mathrm{inter},(I,I+1)} = -J \sigma_{i+1}^{z} \sigma_{i+2}^{z} - h_{x} \sigma_{i+1}^{x} - h_{z} \sigma_{i+1}^{z} \label{eq:Hinter1D},
\end{gather}
where spins $i$ and $i+1$ belong to block $I$, and spin $i+2$ belongs to block $I+1$.
Most importantly,  this particular block partition is suited to preserve the form of the Hamiltonian under the renormalization-group transformations.

The eigenvalues of $H_{\mathrm{intra}}$ are 
\begin{gather}
\varepsilon_{1} = - \sqrt{\left(J + h_{z} \right)^{2} +h_{x}^{2}}, \ \ \ 
\varepsilon_{2} = - \sqrt{\left(J - h_{z} \right)^{2} +h_{x}^{2}}, \label{eq:eigenvalue1} \\
\varepsilon_{3} = \sqrt{\left(J - h_{z} \right)^{2} +h_{x}^{2}}, \ \ \ 
\varepsilon_{4} = \sqrt{\left(J + h_{z} \right)^{2} +h_{x}^{2}}.
\end{gather}
The corresponding eigenvectors are
\begin{gather}
|1\rangle = a_{1,1} |\uparrow \uparrow \rangle + a_{-1,1} |\downarrow \uparrow \rangle, \ \ \ 
|2\rangle = a_{1,-1} |\downarrow \downarrow \rangle + a_{-1,-1} |\uparrow \downarrow \rangle, \\
|3\rangle = a_{-1,-1} |\downarrow \downarrow \rangle - a_{1,-1} |\uparrow \downarrow \rangle, \ \ \ 
|4 \rangle = a_{-1,1} |\uparrow \uparrow \rangle - a_{1,1} |\downarrow \uparrow \rangle,
\end{gather}
where
\begin{equation}
a_{b,c} = \sqrt{\frac{1}{2} \left( 1 + b \, \frac{J + c \, h_{z}}{\sqrt{\left( J + c \, h_{z} \right)^{2} + h_{x}^{2}}} \right)}, \label{eq:a}
\end{equation}
and $\{|\uparrow \uparrow \rangle, |\uparrow \downarrow \rangle, |\downarrow \uparrow \rangle, |\downarrow \downarrow \rangle \}$ is the orthonormal basis in the $\sigma^z$-basis, i.e.  
$\sigma^{z}|\uparrow\rangle=|\uparrow\rangle$, 
$\sigma^{z}|\downarrow\rangle=-|\downarrow\rangle$.

We next keep the two lowest lying energy eigenstates $|1 \rangle$ and $|2 \rangle$, 
and drop the others, $|3 \rangle$ and $|4 \rangle$.
This procedure is expected to be effective for the study of the ground state.
We then replace each block with a single spin representing the $|1 \rangle$ and $|2 \rangle$ states.
To this end, we define the projector onto the coarse-grained system as
\begin{equation}
P = \bigotimes_{I = 1}^{N/2} P_I,
\end{equation}
where $P_{I}$ is the projector,
\begin{equation}
P_{I} = \left( |1\rangle \langle1| + |2\rangle \langle2| \right)_{I}.
\end{equation}
The resulting coarse-grained Hamiltonian is $PHP$, and the renormalized intra-block Hamiltonian is represented as
\begin{equation}
  P_{I} H_{\mathrm{intra},I} P_{I} = \frac{1}{2} \left( \varepsilon_{1} + \varepsilon_{2} \right) 1_{I} + \frac{1}{2} \left( \varepsilon_{1} - \varepsilon_{2} \right) \sigma_{I}^{z}.
\end{equation}
The corresponding projection of the terms in the inter-block Hamiltonian is written as
\begin{equation}
  P_{I} \left(1_{i} \otimes \sigma_{i+1}^{z} \right) P_{I} = \sigma_{I}^{z}, \label{eq:rsz}
\end{equation}
\begin{equation}
  \begin{split}
  P_{I+1} \left( \sigma_{i+2}^{z} \otimes 1_{i+3} \right) P_{I+1} = \frac{1}{2} & \left(a_{1,1}^2  -a_{1,-1}^2 - a_{-1,1}^2 + a_{-1,-1}^2 \right) 1_{I+1} \\ & + \frac{1}{2} \left( a_{1,1}^2 + a_{1,-1}^2 - a_{-1,1}^2 - a_{-1,-1}^2 \right) \sigma_{I+1}^{z}, \label{eq:lsz}
  \end{split}
\end{equation}
\begin{equation}
  P_{I} \left( 1_{i} \otimes \sigma_{i+1}^{x} \right) P_{I} = \left( a_{1,1} a_{-1,-1} + a_{1,-1} a_{-1,1} \right) \sigma_{I}^{x}. \label{eq:rsx}
\end{equation}
The renormalized Hamiltonian is expressed as
\begin{equation}
  \begin{split}
  PHP &= \sum_{I=1}^{N/2} \frac{\varepsilon_{1} + \varepsilon_{2}}{2} 1_{I} - \tilde{J} \sum_{I=1}^{N/2-1} \sigma_{I}^{z} \sigma_{I+1}^{z} - \tilde{h}_{x} \sum_{I=1}^{N/2} \sigma_{I}^{x} - \tilde{h}_{z} \sum_{I=1}^{N/2} \sigma_{I}^{z}, \label{eq:1DPHP}
  \end{split}
\end{equation}
where
\begin{equation}
\tilde{J} = \frac{J}{2} \left( a_{1,1}^2 + a_{1,-1}^2 - a_{-1,1}^2 - a_{-1,-1}^2 \right),
\end{equation}
\begin{equation}
\tilde{h}_{x} = h_{x} \left( a_{1,1} a_{-1,-1} + a_{1,-1} a_{-1,1} \right),
\end{equation}
\begin{equation}
\tilde{h}_{z} = - \frac{1}{2} \left( \varepsilon_{1} - \varepsilon_{2} \right) + \frac{J}{2} \left( a_{1,1}^2 - a_{1,-1}^2 - a_{-1,1}^2 + a_{-1,-1}^2 \right) +h_{z}. \label{eq:1Dtildehz}
\end{equation}
Note that our transformation preserves the form of the Hamiltonian.
Other choices of the intra- and inter-block Hamiltonians lead to more inconvenient transformations that do not preserve the form of the Hamiltonian. 
In other words, our method does not generate additional coupling constants under renormalization.

Equations (\ref{eq:1DPHP})-(\ref{eq:1Dtildehz}) have a non-trivial fixed point at $\left( k_{x}, k_{z} \right) = \left( 1, 0 \right)$, where $k_{x}$ denotes $h_{x}/J$ and $k_{z}$ is for $h_{z}/J$, which is the exact critical point of the transverse-field Ising model in the absence of longitudinal fields.
Keeping the Hamiltonian for $h_z=0$ self-dual through the transformations would 
account for this outstanding result.
We can also calculate the eigenvalues of the linearized renormalization group transformation, 
and determine the critical exponents $\nu$ and $\eta$.
The results are shown in Table \ref{table:1DIsing}.
\begin{table}
\caption{\label{table:1DIsing}The transition point $k_{c,x}$, the exponents $y_{x}$ (related to $h_{x}$) and $y_{z}$ (related to $h_{z}$) of the linearized renormalization group transformation for the one-dimensional
transverse-field Ising model.
The critical exponents $\nu$ and $\eta$ are determined from $\nu = 1/y_{x}$ and $\eta = d - 2y_{z} + 2$, where $d \ (= 2)$ is the spatial dimension of the corresponding classical system.
The exact solution is for the two-dimensional classical Ising model, which is equivalent (through a quantum-classical mapping~\cite{Nishimori}) to the one-dimensional transverse-field Ising model.}
\begin{ruledtabular}
\begin{tabular}{cccccccc}
 & $k_{c,x}$ & $y_{x}$ & $y_{z}$ & $\nu$ & $\eta$ \\
\hline
exact solution & $1$ & $1$ & $1.875$ & $1$ & $0.25$  \\ 
renormalization group & $1$ & $1$ & $1.543$ & $1$ & $0.914$ \\
\end{tabular}
\end{ruledtabular}
\end{table}
It is remarkable that the exponent $\nu$ is exact, since real-space renormalization group calculations seldom yield exact results.
However, the other exponent $\eta$, which is related to the longitudinal field, is not exact.
The longitudinal field disturbs the spectral symmetry essential to obtain exact values for the critical exponents.
It is natural to think that the key to obtain the exact transition point is the existence of self-duality, while the 
property essential for an accurate determination of critical exponents may not always coincide with the self-dual 
character of the problem.
The construction of block Hamiltonians described above leads to a good estimation of the critical exponent $\nu$ and preserves the self-duality in one dimension when $h_z=0$.
The same construction is effective also in the higher-dimensional transverse-field Ising model, which is not self-dual.
This fact is discussed in the subsequent sections.

\section{\label{sec:2D}Generalization to two dimensions}

\begin{figure}
\begin{center}
\includegraphics[width=4cm,clip]{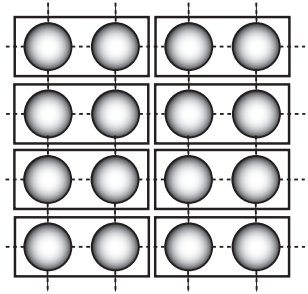}
\caption{Construction of block spins in two dimensions.
This block partition preserves the form of the Hamiltonian under renormalization group transformations.}
\label{fig:2Dblocks}
\end{center}
\end{figure}

We next generalize the method to the two-dimensional transverse-field Ising model on the square lattice.
For simplicity, we set $h_{z} = 0$, and the Hamiltonian becomes
\begin{equation}
H = -J \sum_{\langle i, j \rangle} \sigma_{i}^{z} \sigma_{j}^{z} - h_{x} \sum_{i} \sigma_{i}^{x},
\end{equation}
where spins interact with their nearest neighbors $\langle i, j \rangle$.
The key idea consists of performing renormalization group transformations which preserve the form of the Hamiltonian by a projective isometry that preserves the bond algebra (i.e. the algebra realized by the operators $\sigma_{i}^{z} \sigma_{j}^{z}$ and $\sigma_{i}^{x}$).
The method to construct blocks and block Hamiltonians is especially crucial. 
Exploiting our experience in one dimension, we divide the lattice into blocks just as in one dimension (Fig.~\ref{fig:2Dblocks}).
Furthermore, we combine the one-dimensional block method in horizontal and vertical directions 
to restore the symmetry of the lattice:
If we iterate the renormalization in the same way as in one dimension, the system will be rescaled in only one direction.
To renormalize the system also in the other direction, we iterate the renormalization in two directions; 
first in the horizontal direction and then in the vertical direction.

In the first step of the renormalization (in the horizontal direction) we can replace each block with a single spin using the same procedure as in the one-dimensional case.
For $h_{z} = 0$, Eqs. (\ref{eq:eigenvalue1}) and (\ref{eq:a}) are
\begin{equation}
\varepsilon_{1} = \varepsilon_{2} = - \sqrt{J^{2} + h_{x}^{2}}, 
\end{equation}
\begin{equation}
a_{b,1} = a_{b,-1} = \sqrt{\frac{1}{2} \left( 1 + b \, \frac{J}{\sqrt{J^{2} + h_{x}^{2}}} \right)},
\end{equation}
and the equations corresponding to Eqs. (\ref{eq:rsz}), (\ref{eq:lsz}) and (\ref{eq:rsx}) are
\begin{equation}
P_{I} \left(1_{i} \otimes \sigma_{i+1}^{z} \right) P_{I} = \sigma_{I}^{z},
\end{equation}
\begin{equation}
P_{I+1} \left( \sigma_{i+2}^{z} \otimes 1_{i+3} \right) P_{I+1} = \frac{J}{\sqrt{J^{2} + h_{x}^{2}}} \sigma_{I+1}^{z},
\end{equation}
\begin{equation}
P_{I} \left( 1_{i} \otimes \sigma_{i+1}^{x} \right) P_{I} = \frac{h_{x}}{\sqrt{J^{2} + h_{x}^{2}}} \sigma_{I}^{x}.
\end{equation}
We find that the $z$-component of the spin on the right spot in a block becomes the $z$-component of the block spin, but the $z$-component of the spin on the left spot in a block becomes the $z$-component of the block spin multiplied by $J/ \sqrt{J^{2} + h_{x}^{2}}$.
Now, let us redefine the coupling constants for the horizontal direction and the vertical direction as $J_{h}$ and $J_{v}$ in order to distinguish these two quantities in this scheme.
The renormalized coupling constants and transverse field are then written as,
\begin{equation}
\tilde{J}_{h} = \frac{J_{h}^{2}}{\sqrt{J_{h}^{2} + h_{x}^{2}}},
\end{equation}
\begin{equation}
\tilde{J}_{v} = J_{v} \left( \frac{J_{h}^2}{J_{h}^{2} + h_{x}^{2}} +1 \right), \label{eq:tildeJ}
\end{equation}
\begin{equation}
\tilde{h}_{x} = \frac{h_{x}^{2}}{\sqrt{J_{h}^{2} + h_{x}^{2}}}.
\end{equation}
In Eq. (\ref{eq:tildeJ}), $J_{v} J_{h}^{2}/(J_{h}^{2} + h_{x}^{2})$ is derived from the coupling of two spins on the left spot in each block, and the rest is derived from the one on the right spot in the blocks.

Next the system is renormalized in the vertical direction in the same way as the horizontal direction to recover the symmetry. The coupling constants and the transverse field are now
\begin{equation}
\tilde{\tilde{J}}_{h} = \tilde{J}_{h} \left( \frac{\tilde{J}_{v}^2}{\tilde{J}_{v}^{2} + \tilde{h}_{x}^{2}} +1 \right),
\end{equation}
\begin{equation}
\tilde{\tilde{J}}_{v} = \frac{\tilde{J}_{v}^{2}}{\sqrt{\tilde{J}_{v}^{2} + \tilde{h}_{x}^{2}}},
\end{equation}
\begin{equation}
\tilde{\tilde{h}}_{x} = \frac{\tilde{h}_{x}^{2}}{\sqrt{\tilde{J}_{v}^{2} + \tilde{h}_{x}^{2}}}.
\end{equation}
{}From these equations, the following renormalization group equations are generated,
\begin{equation}
\tilde{\tilde{k}}_{h} = \frac{k_{h}^3 k_{v} \sqrt{\left( 1+k_{h}^{2} \right) \{ \left(2 + k_{h}^{2} \right)^{2} + k_{h}^{2} k_{v}^{2} \left( 1 + k_{h}^{2} \right) \} }}{2 \left( 2 + k_{h}^{2} \right)^{2} + k_{h}^{2} k_{v}^{2} \left( 1 + k_{h}^{2} \right)},
\end{equation}
\begin{equation}
\tilde{\tilde{k}}_{v} = \frac{k_{h}^{2} k_{v}^{2} \left( 1 + k_{h}^{2} \right)}{\left( 2 + k_{h}^{2} \right)^{2}}.
\end{equation}
These equations can be represented as
\begin{equation}
\tilde{\tilde{k}}_{h}^{2} = f \left( k_{h}^{2}, \tilde{\tilde{k}}_{v} \right), \label{eq:k_hA}
\end{equation}
\begin{equation}
\tilde{\tilde{k}}_{v} = f \left( k_{v}, k_{h}^{2} \right), \label{eq:k_vA}
\end{equation}
\begin{equation}
f\left( x, y \right) = x^{2} \frac{y (1 + y)}{(2 + y)^{2}}.
\end{equation}

\begin{figure}
\begin{center}
\includegraphics[width=9cm,clip]{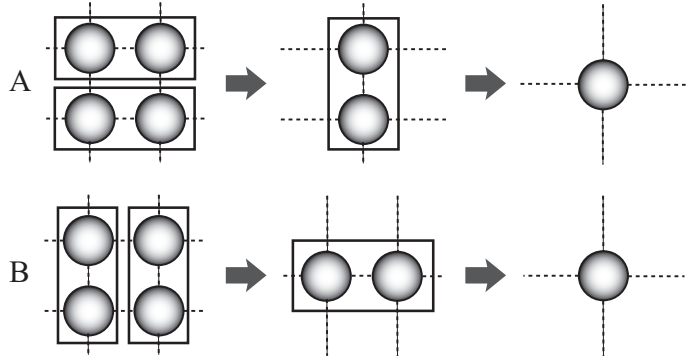}
\caption{The order of the renormalization in the horizontal and vertical directions.
With A, we renormalize the system in the horizontal direction and then in the vertical direction.
B has the reverse order.}
\label{fig:order2D}
\end{center}
\end{figure}

These renormalization group equations are still asymmetric in $k_{h}$ and $k_{v}$.
To render the renormalization symmetric we renormalize the system in the reverse order, vertical and then horizontal.
We define the order of the renormalization in the horizontal direction and then in the vertical direction as the order A, and the reverse order as B (Fig. \ref{fig:order2D}).
Hence, the new step is represented as AB.
The renormalization group transformation with order A is established by 
Eqs.~(\ref{eq:k_hA}) and (\ref{eq:k_vA}) while with order B is
\begin{equation}
\tilde{\tilde{k}}_{v}^{2} = f \left( k_{v}^{2}, \tilde{\tilde{k}}_{h} \right), \label{eq:k_vAB}
\end{equation}
\begin{equation}
\tilde{\tilde{k}}_{h} = f \left( k_{h}, k_{v}^{2} \right). \label{eq:k_hAB}
\end{equation}
The transformation with order AB is obtained from the substitution of $\tilde{\tilde{k}}_{h}$, Eq.~(\ref{eq:k_hA}), for $k_{h}$ in Eq.~(\ref{eq:k_hAB}) and $\tilde{\tilde{k}}_{v}$, Eq.~(\ref{eq:k_vA}), for $k_{v}$ in Eqs.~(\ref{eq:k_vAB}) and (\ref{eq:k_hAB}).
Although B relaxes the asymmetry in A, the renormalization group equations with AB is still asymmetric.
The symmetrization procedure is therefore repeated as ABBA and ABBABAAB.

If we regard the renormalization map in the order ABBABAAB of scaling factor $2^8$ as a single transformation, the eigenvalues of the linearized renormalization group transformation are $7731.18 \ (=(2^{8})^{1.61456})$ and $8.477 \ ( = (2^{8})^{0.385441} )$.
One of the eigenvalues is much larger than the other, and hence we may be justified to ignore the smaller eigenvalue.
The value of the critical exponent $\nu$ derived from the larger eigenvalue is $0.61936$, which is very close to the reliable numerical result $0.6301$~\cite{Pelissetto}. 

\begin{figure}
\begin{center}
\includegraphics[width=11cm,clip]{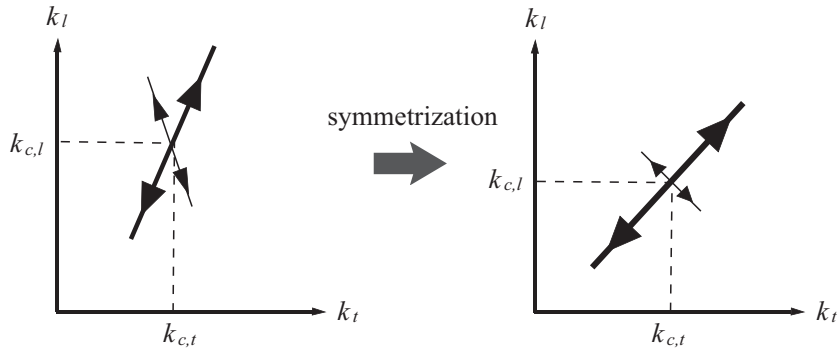}
\caption{The renormalization group flow near the fixed point in the space of $k_{h}$ and $k_{v}$.
With the symmetrization procedure $k_{c,h}$ and $k_{c,v}$ become close to each other, and the slope of the dominant flow (the thicker line) approaches $1$, and the subdominant direction becomes perpendicular.}
\label{fig:RG flow}
\end{center}
\end{figure}

The effectiveness of the symmetrization scheme is clearly seen in Table~\ref{table:2D}.
\begin{table}
\caption{\label{table:2D}The critical point $k_{c}$, eigenvalues of the renormalization group transformation $\lambda$, critical exponent $\nu$, the slope of the dominant eigenvector $\boldsymbol{\phi_{1}}$, and the scalar product of two eigenvectors $\boldsymbol{\phi_{1}} \cdot \boldsymbol{\phi_{2}}$ from real-space renormalization group with symmetrization on the square lattice.} 
\begin{ruledtabular}
\begin{tabular}{cccccc}
 & A & AB & ABBA & ABBABAAB & three-dimensional classical \\
\hline
$k_{c,h}$ & $1.544$ & $1.793$ & $1.828$ & $1.835$ & $3.4351$~\cite{Blote} \\ 
$k_{c,v}$ & $2.383$ & $1.897$ & $1.855$ & $1.848$ & \\
$\lambda_{1}$ & $3.083$ & $9.377$ & $87.901$ & $7731.18$ &  \\ 
$\lambda_{2}$ & $1.297$ & $1.706$ & $2.912$ & $8.477$ &  \\ 
$\nu_1$ & $0.61555$ & $0.61935$ & $0.61941$ & $0.61936$ & $0.6301(4)$~\cite{Pelissetto} \\ 
$\nu_2$ & $2.66356$ & $2.59466$ & $2.59372$ & $2.59443$ & \\
slope of $\boldsymbol{\phi_{1}}$ & $1.758$ & $1.059$ & $1.026$ & $1.002$ & \\
$\boldsymbol{\phi_{1}} \cdot \boldsymbol{\phi_{2}}$ & $0.644$ & $0.140$ & $0.071$ & $0.050$ & \\
\end{tabular}
\end{ruledtabular}
\end{table}
With only A, $k_{c,h}$ and $k_{c,v}$ are very different, and both eigenvalues of the linearized renormalization group transformation are relevant.
The introduction of reverse order relaxes the asymmetry of the fixed point.
In addition, in the space of $k_{h}$ and $k_{v}$, the slope of the dominant flow approaches $1$, and the other flow becomes perpendicular to the dominant one (Fig.~\ref{fig:RG flow}).
These changes suggest that the larger eigenvalue is the reliable one and we may ignore the other.
The latter seems to be an artifact of the approximation.

We can also study the case with a longitudinal field, $h_{z} \neq 0$.
With the same scheme as for $h_{z} = 0$, we obtain the eigenvalues of the linearized renormalization group transformation and critical exponents as listed in Table \ref{table:2Dc.e}.
\begin{table}
\caption{\label{table:2Dc.e}The exponents $y_{x}$ and $y_{z}$ for the linearized renormalization group transformation and the critical exponents $\nu$ and $\eta$ for the square lattice, derived from symmetrization.}
\begin{ruledtabular}
\begin{tabular}{cccccc}
\ \ \ \ \ \ \ \ \ \ & A & AB & ABBA & ABBABAAB & three-dimensional classical~\cite{Pelissetto} \\
\hline
$y_{x}$ & $1.62456$ & $1.61459$ & $1.61445$ & $1.61456$ &  \\ 
$y_{z}$ & $2.39774$ & $2.38895$ & - & - &  \\
$\nu$ & $0.61555$ & $0.61935$ & $0.61941$ & $0.61936$ & $0.6301(4)$ \\
$\eta$ & $0.20452$ & $0.22209$ & - & - & $0.0364(5)$ \\
\end{tabular}
\end{ruledtabular}
\end{table}
The exponent $y_{z}$, or the critical exponent $\eta$, is not very close to the numerical result of the classical Ising model in three dimensions~\cite{Pelissetto}, a
situation similar to the one-dimensional case.
Moreover, this value is not improved by the symmetrization process.
This property may give an important clue for the study of the relation between the symmetry of a model and the symmetry of a method.

\section{\label{sec:3D}three dimensions}

\begin{figure}
\begin{center}
\includegraphics[width=16cm,clip]{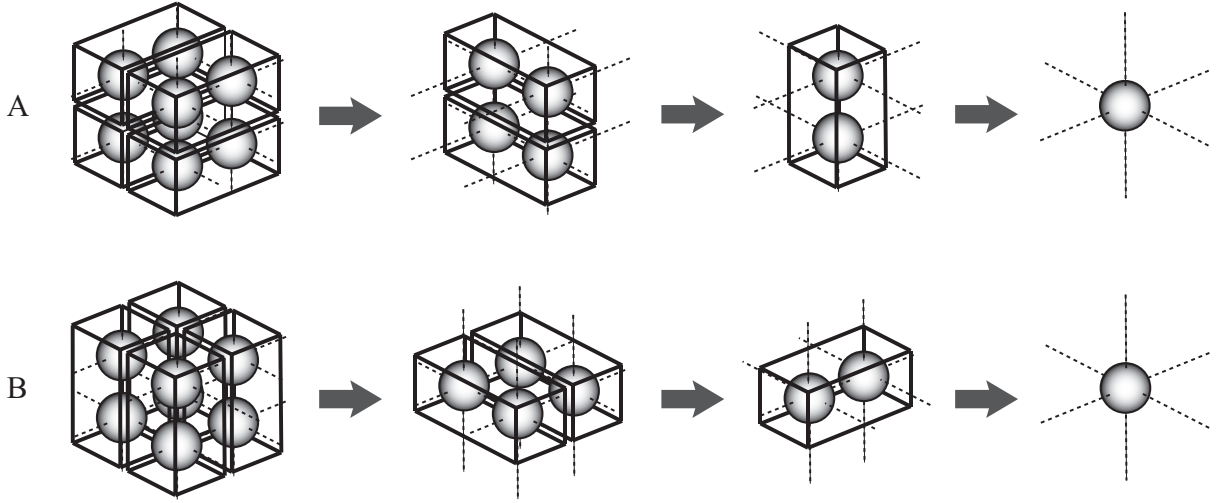}
\caption{The order of renormalization in three dimensions.
With A, we renormalize the system in the horizontal, then vertical  and finally in the third direction.
B realizes the reverse order.}
\label{fig:order3D}
\end{center}
\end{figure}

Next, we generalize the scheme to a cubic lattice in three dimensions.
Let us again restrict ourselves to the system without longitudinal field for simplicity.
We have to renormalize the three-dimensional system in three directions.
The order A is defined as horizontal, then vertical and finally along the third direction.
The order B is the reverse of A (Fig.~\ref{fig:order3D}). 
We now define the coupling constant for the third direction as $J_{t}$ in addition to $J_{h}$ and 
$J_{v}$ for the horizontal and vertical directions.

In the first step of order A, which is the renormalization in the horizontal direction, the parameters change as 
\begin{equation}
\tilde{J}_{h} = \frac{J_{h}^{2}}{\sqrt{J_{h}^{2} + h_{x}^{2}}},
\end{equation}
\begin{equation}
\tilde{J}_{v} = J_{v} \left( \frac{J_{h}^2}{J_{h}^{2} + h_{x}^{2}} +1 \right), 
\end{equation}
\begin{equation}
\tilde{J}_{t} = J_{t} \left( \frac{J_{h}^2}{J_{h}^{2} + h_{x}^{2}} +1 \right), 
\end{equation}
\begin{equation}
\tilde{h}_{x} = \frac{h_{x}^{2}}{\sqrt{J_{h}^{2} + h_{x}^{2}}}.
\end{equation}
Note that the coupling constants of vertical and third directions are changed under the same rule.
In general, when the system is renormalized in the direction $\alpha$ and another direction is represented 
as $\beta$, the parameters are transformed as 
\begin{equation}
\tilde{J}_{\alpha} = \frac{J_{\alpha}^{2}}{\sqrt{J_{\alpha}^{2} + h_{x}^{2}}},
\end{equation}
\begin{equation}
\tilde{J}_{\beta} = J_{\beta} \left( \frac{J_{\alpha}^2}{J_{\alpha}^{2} + h_{x}^{2}} +1 \right), 
\end{equation}
\begin{equation}
\tilde{h}_{x} = \frac{h_{x}^{2}}{\sqrt{J_{\alpha}^{2} + h_{x}^{2}}}.
\end{equation}
Carrying out the three steps of the scheme with these relations, we can obtain the parameters of 
the system renormalized in the three directions.

The symmetrization with the combinations of A and B as in the two-dimensional system improves the result (Table ~\ref{table:3Dc.e}).
\begin{table}
\caption{\label{table:3Dc.e} The exponents $y_{x}$ and $y_{z}$ for the linearized renormalization group 
transformation and the critical exponents $\nu$ and $\eta$ for the cubic lattice, derived from symmetrization.}
\begin{ruledtabular}
\begin{tabular}{cccc}
\ \ \ \ \ \ \ \ \ \ & A & AB & classical mean-field \\
\hline
$y_{x}$ & $2.0213$ & $2.0092$ & $2$  \\ 
$y_{z}$ & $3.1507$ & $3.1428$ & $3$ \\
$\nu$ & $0.49474$ & $0.49772$ & $\frac{1}{2}$ \\
$\eta$ & $-0.30145$ & $-0.28553$ & $0$ \\
\end{tabular}
\end{ruledtabular}
\end{table}
The value of $\nu$ with A is $0.49474$ and with AB is $0.49772$.
In the classical mean-field Ising model, which corresponds to the three-dimensional transverse-field 
Ising model, the value of $\nu$ is $1/2$.
Our procedure yields almost the exact value of $\nu$, and the symmetrization is an effective way to 
improve the results just as in the two-dimensional case.

The result of the exponent related to a longitudinal field in three dimensions is also shown in Table ~\ref{table:3Dc.e}.
The value of $\eta$ is not as good as the result for $\nu$, a situation similar to the
one- and two-dimensional cases.

\section{\label{sec:conclusions}Conclusions}

In this paper, we proposed a real-space renormalization group procedure for the transverse-field Ising model in finite dimensions.
The procedure is based on the block-spin transformation, and the preservation of the form of the Hamiltonian under the transformation is essential to successfully extract the critical point and the critical exponents.
A remarkable feature of the method is that it yields exact results in one dimension~\cite{Fernandez}.
We have generalized this idea to higher dimensions.
The one-dimensional block method is also effective in higher dimensions, and we have combined the method in 
horizontal, vertical and the third directions to restore the symmetry of the lattice.
Our results demonstrate the utility of the block Hamiltonian we have adopted.
Although the results fall short of the exact solutions as in the one-dimensional case, they still represent important steps because the real-space renormalization group rarely gives accurate estimates of critical exponents in three and four dimensions.

There are several points to be clarified concerning the present method.
We are particularly interested in establishing the reasons why the one-dimensional case 
yields exact results since
the answer may give an important hint on how to improve the higher-dimensional cases.
The exact result in one dimension strongly suggests that one of the key points 
is the fact that the one-dimensional transverse-field Ising model is self-dual.
Since the critical exponent $\nu$ has been estimated to good accuracy in two and three dimensions 
where there is no self-duality, additional factors should have contributed such as the preservation 
of the bond algebra embedded in our construction of the block Hamiltonian.
It is necessary to clarify what has been the essential ingredient for further developments.
It is also a topic of interest to generalize our technique to other systems including the Potts model and 
disordered systems.
The latter is important due to the lack of reliable analytical approaches to three-dimensional systems.

\begin{acknowledgments}
This work was partially supported by CREST, JST.
\end{acknowledgments}

\bibliography{basename of .bib file}

\end{document}